\documentclass[twocolumn,prl,groupedaddress,preprintnumbers,amsmath,amssymb,showpacs,floatfix]{revtex4}

\usepackage{graphicx}
\usepackage{dcolumn}
\usepackage{bm}

\begin{document}

\preprint{Phys. Rev. Lett. 81, 922--925, (1998)}

\title{Nonequilibrium Electron Interactions in Metal Films}

\author{N. Del Fatti}
\author{R. Bouffanais}%
\author{F. Vall\'ee}
\author{C. Flytzanis}
\affiliation{%
Laboratoire d'Optique Quantique du CNRS, Ecole Polytechnique, 91128 Palaiseau Cedex, France
}%

\date{Received 4 February 1998}

\begin{abstract}
Ultrafast relaxation dynamics of an athermal electron distribution is investigated in silver films using a femtosecond pump-probe technique with 18 fs pulses in off-resonant conditions. The results yield evidence for an increase with time of the electron-gas energy loss rate to the lattice and of the free electron damping during the early stages of the electron-gas thermalization. These effects are attributed to transient alterations of the electron average scattering processes due to the athermal nature of the electron gas, in agreement with numerical simulations.
\end{abstract}

\maketitle
The average properties of a free electron gas, such as the average electron scattering rate and the electron-gas energy losses, play a central role in the fundamental and technological properties of metallic and semiconductor systems. Although these are basically determined by the individual electron interactions, a collective behavior, related to the electron temperature, is observed for a thermalized electron gas. A drastically different behavior is expected for an athermal electron gas with thus strongly modified average electron properties.

With the advance of femtosecond lasers, such transient athermal electron distribution can now be created and probed. In particular, in the case of noble metal systems, it has been shown that screening of the Coulomb interaction and the Pauli exclusion principle effect drastically reduce the electron-electron scattering probability, leading to a slow internal thermalization of the electron gas on a few hundred femtosecond time scale \cite{ref1,ref2,ref3,ref4}. This longlasting nonequilibrium situation offers the unique possibility of analyzing the interactions of an athermal free electron gas with its environment and their dynamic evolution during electron internal thermalization.

Using silver as a model system, we have investigated the properties of a nonequilibrium electron gas using a femtosecond pump-probe technique. We show that off-resonant probing (i.e., far from the interband transitions) of the optical property changes of a metal film after ultrafast perturbation of its electron distribution, permits selective investigation of the free electron-gas energy loss rate to the lattice and of the average electron scattering rate. The results yield evidence for an increase of both of these rates as the initially athermal electron gas evolves to a thermalized distribution, due to the interplay between the different scattering processes for strongly nonequilibrium electron distributions. 

\begin{figure}[ht!]
\includegraphics[width=0.33\textwidth]{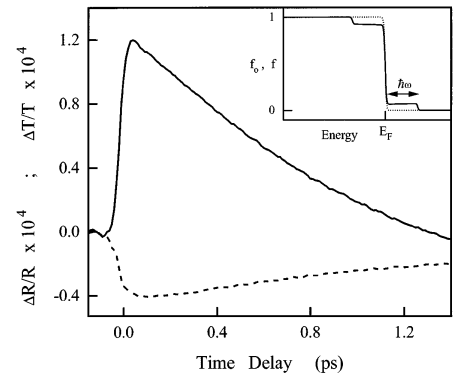}
\caption{\footnotesize Transient transmission (solid line) and reflection (dashed line) changes measured in a 20 nm thick Ag film for infrared excitation and probing ($\hbar\omega \sim $ 1.45 eV) at $T_0=295$ K. The pump fluence is 190 $\mu$J/cm$^2$. The inset shows the equilibrium electron distribution ($f_0$, dotted line) and the perturbed one ($f$, solid line); $E_F$ indicates the Fermi energy.}\label{fig1}
\end{figure}

Measurements were performed in silver using a femtosecond pump-probe technique with near-infrared pulses to satisfy the off-resonant conditions (the threshold for interband transitions in silver is $\hbar \Omega_{ib} \sim 4$ eV \cite{ref5}). The electron distribution of an optically thin film is perturbed by intraband (free electron) absorption of the transform limited 18 fs infrared pulses delivered by a~Ti$:$sapphire oscillator ($\hbar \omega \sim$ 1.45 eV). Electrons are thus excited from below to above the Fermi energy, creating a strongly athermal distribution (Fig.~\ref{fig1}, inset) that internally thermalizes by electron-electron scattering on a 500 fs time scale \cite{ref1,ref3,ref6}. The induced reflectivity, $\Delta R/R$, and transmissivity, $\Delta T/T$, changes are measured at the same wavelength using a time-delayed probe pulse. This off-resonant probing permits measurements of weak features associated with changes of the electron-gas average properties.  

A conventional pump-probe setup has been used with differential and lock-in detection of the probe beam transmission and reflection changes. The two cross-polarized beams are focused over a focal spot of 30 mm in diameter on a 20 nm thick polycrystalline silver film grown on a sapphire substrate. The high repetition rate (80 MHz) and stability of our femtosecond system permits noise levels for $\Delta T/T$ and $\Delta R/R$ measurements in the $10^{-6}$ range. 
\begin{figure}
\includegraphics[width=0.33\textwidth]{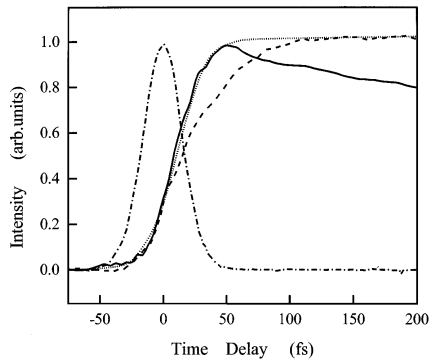}
\caption{\footnotesize Short-time~behavior~of~the~normalized~$\Delta T/T$~(solid line)~and~$\Delta R/R$~(dashed~line).~Also~shown~are~the~pump-probe cross~correlation~(dash-dotted~line)~and~its~integral~(dotted~line).}\label{fig2}
\end{figure}

The measured transient reflectivity, $\Delta R/R$, and transmissivity, $\Delta T/T$, changes are shown in Fig.~\ref{fig1} for a~pump fluence of 190 $\mu$J/cm$^2$ that would correspond to a maximum electron temperature rise $\Delta T_e \sim 90$ K for a thermalized electron gas. Similar results were obtained for pump fluences in the range $500-4$ $\mu$J/cm$^2$ (i.e., $\Delta T_e \sim  200-2$ K). The observed changes show very different short time delay behaviors. The $\Delta T/T$ rise closely follows energy injection in the electron gas [i.e., the integral of the pump-probe correlation (Fig.~\ref{fig2})]. In contrast, $\Delta R/R$ exhibits a delayed rise and reaches its maximum value after about 120 fs, indicating that $\Delta R/R$ and $\Delta T/T$ are sensitive to different electronic property changes. No modification of the measured responses was observed with silver film aging, indicating that oxidation does not significantly influence our results. 

\begin{figure}
\includegraphics[width=0.33\textwidth]{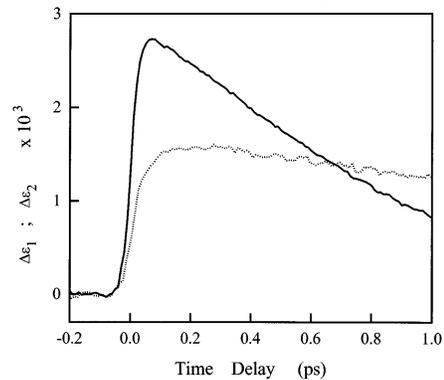}
\caption{\footnotesize Induced changes of the real part ($\Delta \varepsilon_1$, solid line) and imaginary part ($\Delta \varepsilon_2$, dotted line) of the metal dielectric constant obtained from the measured transmission and reflection changes in the 20-nm-thick film.}\label{fig3}
\end{figure}

In the weak perturbation regime investigated here, $\Delta R/R$ and $\Delta T/T$ are linear combinations of the changes of the real, $\Delta \varepsilon_1$, and imaginary, $\Delta \varepsilon_2$, parts of the metal dielectric function. The coefficients of these combinations can be calculated from the known optical properties of silver \cite{ref7} taking into account Fabry-P\'erot effects \cite{ref3}. For our probe wavelength, $\Delta T/T$ essentially reflects $\Delta \varepsilon_1$ (the contribution of $\Delta \varepsilon_1$ to $\Delta T/T$ is about 5 times larger than that of $\Delta \varepsilon_2$) while $\Delta R/R$ reflects both $\Delta \varepsilon_1$ and $\Delta \varepsilon_2$, whose transient behaviors can thus be determined (Fig.~\ref{fig3}). A striking difference here is that, as $\Delta T/T$, $\Delta \varepsilon_1$ exhibits no rise time while, as $\Delta R/R$, $\Delta \varepsilon_2$ reaches its maximum value after about 120 fs. This demonstrates the different physical origins of the induced changes of the dielectric constant dispersive and absorptive parts. 

The metal dielectric constant $\varepsilon=\varepsilon_1 + i \varepsilon_2$ can be written as the sum of free (intraband) and bound (interband) electron contributions: 
\begin{equation}\label{eq1}
\varepsilon(\omega) = \varepsilon^b(\omega ) - \frac{\omega^2_p}{\omega (\omega + i \gamma )}, 
\end{equation}
where $\omega_p$ is the plasmon frequency and $\gamma$ is the free electron optical scattering rate. In noble metals the interband term $\varepsilon^b$ is dominated by transitions from the full $d$ bands to the Fermi surface states. Energy injection into the electron gas results in a spreading of the electron distribution around the Fermi energy, $\hbar\Omega_F$, and thus to an induced (reduced) absorption for interband transitions to electron states below (above) $\hbar\Omega_F$. These transitions, however, take place at large photon energies and hence, for sufficiently small pump and probe photon energy (i.e., $2\hbar \omega < \hbar\Omega_{ib}$), $\Delta \varepsilon_2$ is only due to changes of the intraband term ($\Delta \varepsilon_2^b$). In contrast, $\Delta \varepsilon_1$, which is sensitive to the integrated absorption spectrum, is dominated by changes of the interband term $\Delta \varepsilon_1^b$ (the change of the real part of the free electron contribution $\Delta \varepsilon_1^{\textrm{\scriptsize intra}}$ is negligible; see below). This can be calculated using the Kramers-Kronig relation: 
\begin{widetext}
\begin{equation}\label{eq2}
\Delta \varepsilon_1^b(\omega ) \propto \int \frac{(\omega '-\Omega_F+\Omega_{ib})^{-1}g(\omega ') [f_0 (\hbar \omega ' ) - f (\hbar \omega ' )]}{(\omega ' -\Omega_F+\Omega_{ib})^2-\omega^2} d\omega ',
\end{equation}
\end{widetext}
where undispersed $d$ bands have been assumed. $f_0$ and $f$ are the equilibrium and transient electron distribution functions and $g$ is the conduction band density of states. In off-resonant conditions, one can show that $\Delta \varepsilon_1^b$ is proportional to the electron-gas excess energy, $\Delta E_e$ \cite{ref4}:
\begin{equation}\label{eq3}
\Delta \varepsilon_1^b(\omega ) \propto \int \omega^{\prime 3/2} [f (\hbar \omega ' ) - f_0 (\hbar \omega ' )] d\omega ' \propto \Delta E_e .
\end{equation}
The short time delay increase of $\Delta \varepsilon_1$ thus essentially reflects the time evolution of the excess energy stored in the electron gas and is expected to rise instantaneously, in agreement with the experimental results. Furthermore, its delay yields information on energy transfer to the lattice permitting direct investigation of its dependence on the electron distribution. 

\begin{figure}
\includegraphics[width=0.29\textwidth]{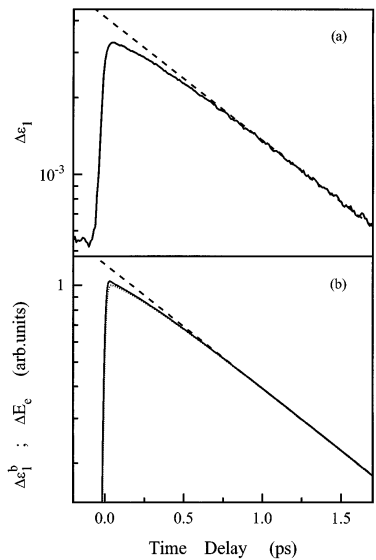}
\caption{\footnotesize (a) Measured transient change of the real part $\Delta \varepsilon_1$ of the dielectric constant on a logarithmic scale. The dashed line corresponds to an exponential decay with $\tau = 900$ fs. (b) Calculated evolution of the real part change of the interband contribution to the dielectric constant, $\Delta \varepsilon_1^b$ (solid line),~and~of the~excess~energy~stored~in~the~electron~gas,~$\Delta E_e$~(dotted~line).}\label{fig4}
\end{figure}

The temporal dependence of $\Delta \varepsilon_1$ is shown in Fig.~\ref{fig4} on a logarithmic scale after subtraction of its final background value. The initial decay of $\Delta \varepsilon_1$ clearly shows that the electron-gas energy losses to the lattice are initially slow and increase over a time scale of a few hundred femtoseconds (Fig.~\ref{fig4}). For longer times, a constant rate is reached, corresponding to an exponential decay (Fig.~\ref{fig4}), in agreement with the theoretical results for a thermalized electron gas (two-temperature model \cite{ref8,ref9}) in the weak perturbation regime \cite{ref3}. The time constant of the long term exponential decay yields an effective electron-phonon coupling time of $\sim$900 fs, consistent with previous measurements \cite{ref4}. A similar behavior has been observed in gold films with, however, an additional fast transient peak due to transient induced interband absorption at the probe photon energy (in contrast to silver, $2\hbar \omega > \hbar \Omega_{ib}$ in gold and $\Delta \varepsilon_2^b(\omega) \neq 0$ for very short time delays). 

This increase with time of the effective electron gas-lattice coupling is a direct consequence of the noninstantaneous internal thermalization of the electron gas. During the excitation process, a very small number of electrons gains a large excess energy as compared to $k_{\textrm{B}}T_0$. As a crude approximation, separating the electron gas into unperturbed and nonequilibrium electrons, only the latter ones can lose energy by phonon emission. Their number increases with time as electron-electron scattering redistributes energy among the carriers: this leads to an overall increase of the energy loss rate to the lattice during the early stages of the internal electron-gas thermalization \cite{ref2,ref4}. As internal thermalization is approached, the above separation is no longer valid and a constant energy loss rate is eventually reached, corresponding to collective electron-gas-lattice interaction as described by the two-temperature model. This evolution from a quasi-individual to a collective electron behavior is responsible for the observed short time nonexponential excess energy decay. 

The above description of the interplay between electron thermalization and energy losses can be made more quantitative by computing the electron-gas relaxation dynamics using the Boltzmann equation for electrons, including electron-electron and electron-phonon scattering. Numerical simulations were performed for a free electron gas using statically screened electron-electron Coulomb interaction \cite{ref3}. Deformation potential electron-phonon interaction has been introduced using the Debye model for the phonon dispersion. The only parameter is the amplitude of the deformation potential coupling which has been set by imposing the computed long time (exponential) signal decay to be identical to the measured one. The time dependent changes of the interband term $\Delta \varepsilon_1^b$ are calculated from the electron distribution using the band structure model of Rosei \emph{et al.} \cite{ref5}. 

The results are in very good agreement with the experimental ones, showing in particular the same transient evolution of the signal decay to an exponential behavior (Fig.~\ref{fig4}). As expected, this behavior is very similar to that of the excess energy stored in the electron gas with only a slight short time delay deviation (Fig.~\ref{fig4}). This is a consequence of the initial perturbation of the occupation number of electron states far below the Fermi energy that leads to more resonant transient contributions to $\varepsilon_1^b$ [Eq.~\eqref{eq2}]. Similar calculations performed assuming instantaneous internal thermalization of the electron gas show only an exponential decay of $\Delta \varepsilon_1^b$, in agreement with the two-temperature model. 

The conduction band electron states probed from the $d$ bands are not perturbed and, in contrast to $\Delta \varepsilon_1$, $\Delta \varepsilon_2 $ reflects only the modification of the intraband optical absorption. Using the Drude model, this is connected to the alteration of the average optical scattering rate $\gamma$ of the electrons: 
\begin{equation}\label{eq4}
\frac{\Delta \varepsilon_2}{\varepsilon_2} \approx \left( \frac{\omega^2-\gamma^2}{\omega^2+\gamma^2}   \right) \frac{\Delta \gamma}{\gamma} \approx \frac{\Delta \gamma}{\gamma}.
\end{equation}
Our results thus demonstrate a noninstantaneous increase of $\gamma$ with energy injection in the electron gas (Fig.~\ref{fig3}), and thus the modification of the efficiency of the scattering processes with the electron distribution character (athermal or thermal). $\Delta \varepsilon_2 $, and thus $\Delta \gamma$, subsequently decays and reaches a plateau with a temporal behavior comparable to that measured for $\Delta \varepsilon_1 $.

For short time delays all of the excess energy is stored in the electron gas, and the increase of $\gamma$ can thus be attributed to changes of the electron scattering process efficiency as the electron distribution broadens in the vicinity of the Fermi energy (weakening of the Pauli exclusion effects). Although at room temperature electron-phonon interaction is the dominant process, the other scattering mechanisms (i.e., electron-defect and umklapp electron-electron scattering) can also contribute to the observed changes. Their computation in the general situation of a nonequilibrium electron distribution is far beyond the scope of this paper, and we have only performed crude estimates using the results for thermal systems \cite{ref10,ref11}. For our probe photon energy, the dominant changes of the optical electron scattering rate is expected to originate from the temperature dependent part of the electron-electron scattering rate \cite{ref10}. For an electron temperature rise of 90 K and a relative contribution of electron-electron scattering to the total scattering rate $\gamma_{e-e}/\gamma \sim 5\times 10^{-2}$ \cite{ref11}, $\Delta \gamma/\gamma$ is estimated to be of the order of $10^{-3}$, corresponding to $\Delta \varepsilon_2\sim 4\times 10^{-4}$ [Eq.~\eqref{eq4} with $\varepsilon_2 \sim 0.5$ \cite{ref7}]. This is consistent with the measured value when one takes into account the approximations in computing umklapp electron-electron scattering and the possible role of the interfaces. The observed $\Delta \varepsilon_2$ rise time suggests that, as for electron-lattice energy exchanges, the electron-electron collision rate also evolves during energy redistribution in the electron gas with its probability increasing on a 100 fs time scale. 

With the lattice heat capacity being much larger than the electron one, as electrons thermalize with the lattice their excess energy strongly decreases, eventually making $\Delta \gamma_{e-e}$ negligible. The concomitant rise of the lattice temperature, $\Delta T_L$, leads to a significant increase of the electron-phonon scattering probability which is proportional to the phonon occupation numbers and thus to $T_L$. This effect eventually dominates the electron scattering rate change, $\gamma_{e-\textrm{\scriptsize ph}}$ finally reaching a constant value determined by $\Delta T_L$: $\Delta \gamma_{e-\textrm{\scriptsize ph}}/\gamma_{e-\textrm{\scriptsize ph}} \sim \Delta T_L/T_L$. Neglecting long time scale heat diffusion, the electron-lattice equilibrium temperature rise is estimated to be $\Delta T_e = \Delta T_L \sim 0.7$ K, corresponding to an estimated long term $\varepsilon_2$ change $\Delta \varepsilon_2\sim 1.1\times 10^{-3}$, in very good agreement with the experimental value (Fig.~\ref{fig3}).
 
Modification of the free electron scattering might also alter the real part of the dielectric function [Eq.~\eqref{eq1}] with a contribution proportional to $\Delta \varepsilon_2$: $\Delta \varepsilon_1^{\textrm{\scriptsize intra}}(\omega) \approx 2\gamma/\omega \Delta \varepsilon_2(\omega)$. With the probe photon frequency being much larger than the electron scattering rate ($\gamma \sim 20$ meV \cite{ref7}), $\Delta \varepsilon_1^{\textrm{\scriptsize intra}}$ estimated from the measured $\Delta \varepsilon_2$ is thus much smaller than the observed $\Delta \varepsilon_1$, indicating a negligible intraband contribution to $\Delta \varepsilon_1$. 

In conclusion, transient optical reflectivity and transmissivity changes have been investigated in silver films using a femtosecond pump-probe technique with 18 fs resolution. Experiments were performed in the weak perturbation regime in off-resonant conditions, permitting one to connect the deduced changes of the real and imaginary parts of the metal dielectric constant to fundamental properties of the free electron gas, i.e., respectively, the electron-gas energy losses and the electron optical scattering rate. 

The results demonstrate that the electron-gas energy losses to the lattice increase with time during the early stages of the electron-gas internal thermalization, in very good agreement with numerical simulations of the free electron dynamics. This behavior reflects evolution from an individual to a collective electron-lattice type of coupling, mediated by electron-electron interactions. Induced free electron optical absorption has also been demonstrated with a rise time of about 100 fs, showing a noninstantaneous increase of the average electron scattering rate. This indicates a strong dependence of the probability of the electron scattering on the electron distribution, and theoretical investigations of these processes for off-equilibrium distributions would be particularly interesting here. The induced scattering is likely dominated by the increase of electron-electron scattering for short time delays and electron-phonon scattering due to the lattice temperature rise as the full system reaches thermal equilibrium.


\begin{thebibliography}{10}
\bibitem{ref1}
W. S. Fann, R. Storz, H. W. K. Tom, and J. Bokor, Phys. Rev. Lett. \textbf{68}, 2834 (1992). 
\bibitem{ref2}
G. Tas and H. J. Maris, Phys. Rev. B \textbf{49}, 15 046 (1994). 
\bibitem{ref3}
 C. K. Sun, F. Vall\'ee, L. H. Acioli, E. P. Ippen, and J. G. Fujimoto, Phys. Rev. B \textbf{50}, 15 337 (1994). 
\bibitem{ref4}
R. H. M. Groeneveld, R. Sprik, and A. Lagendijk, Phys. Rev. B \textbf{51}, 11 433 (1995). 
\bibitem{ref5}
R. Rosei, Phys. Rev. B \textbf{10}, 474 (1974); R. Rosei, C. H. Culp, and J. H. Weaver, Phys. Rev. B \textbf{10}, 484 (1974). 
\bibitem{ref6}
N. Del Fatti, M. Achermann, F. Vall\'ee, and C. Flytzanis (unpublished). 
\bibitem{ref7}
P. B. Johnson and R. W. Christy, Phys. Rev. B \textbf{6}, 4370 (1972). 
\bibitem{ref8}
S. I. Anisimov, B. L. Kapeliovitch, and T. L. Perelman, Sov. Phys. JETP \textbf{39}, 375 (1974). 
\bibitem{ref9}
P. B. Allen, Phys. Rev. Lett. \textbf{59}, 1460 (1987). 
\bibitem{ref10}
R. N. Gurzhi, Sov. Phys. JETP \textbf{35}, 673 (1959). 
\bibitem{ref11}
J. B. Smith and H. Ehrenreich, Phys. Rev. B \textbf{25}, 923 (1982). 
\end{thebibliography}
\end{document}